\documentclass[aps,pre,reprint,groupedaddress,showpacs,preprintnumbers,amsmath,amssymb]{revtex4-1}
\usepackage{graphicx}
\usepackage{dcolumn}
\usepackage{bm}
\usepackage{color}
\begin{document}
\preprint{Submitted, 15.07.2015}
\title{A master equation for force distributions in soft particle packings
- Irreversible mechanical responses to isotropic compression and decompression}
\author{Kuniyasu Saitoh}
\affiliation{Faculty of Engineering Technology, MESA+, University of Twente, Drienerlolaan 5, 7522 NB, Enschede, The Netherlands}
\author{Vanessa Magnanimo}
\affiliation{Faculty of Engineering Technology, MESA+, University of Twente, Drienerlolaan 5, 7522 NB, Enschede, The Netherlands}
\author{Stefan Luding}
\affiliation{Faculty of Engineering Technology, MESA+, University of Twente, Drienerlolaan 5, 7522 NB, Enschede, The Netherlands}
\date{\today}
\begin{abstract}
Mechanical responses of soft particle packings to quasi-static deformations are determined by the microscopic restructuring of force-chain networks,
where complex non-affine displacements of constituent particles cause the irreversible macroscopic behavior.
Recently, we have proposed a master equation for the probability distribution functions of contact forces and interparticle gaps
[K. Saitoh et al.,\ \textit{Soft Matter} \textbf{11},\ 1253 (2015)], where mutual exchanges of contacts and interparticle gaps,\ i.e.\
opening and closing contacts, are also involved in the stochastic description with the aid of Delaunay triangulations.
We describe full details of the master equation and numerically investigate irreversible mechanical responses of soft particle packings to cyclic loading.
The irreversibility observed in molecular dynamics simulations is well reproduced by the master equation if the system undergoes quasi-static deformations.
We also confirm that the degree of irreversible responses is a decreasing function of the area fraction and the number of cycles.
\end{abstract}
\maketitle
\section{Introduction}\label{sec:intro}
%
%
\emph{Soft particle packings},\ e.g.\ colloids, emulsions, foams, glasses, and granular materials, are ubiquitous in nature
and a better understanding of their mechanical properties is crucial for industry and science \cite{lemaitre}.
Different from usual solids, their constituents are \emph{macroscopic particles} such that thermal fluctuations are negligibly small for the individual motions
(except for glasses \cite{glass-review}) and thus their macroscopic behavior purely originates from the mechanics of constituent particles \cite{am0}.
Apart from some crystalline systems,\ e.g.\ colloidal crystals \cite{ccrystal}, their configurations are mostly random (or in amorphous states) 
so that they are temporarily at rest, in mechanical equilibrium, once the system has relaxed to a static state \cite{pp0,pp1,pp2,pp3,pp4,pp5,pp6}.
%
Then, a \emph{packing fraction},\ $\phi$,\ has been used as a measure of rigidity of soft particle packings.
Some physical quantities responsible for either mechanical properties or microscopic structure exhibit the critical behavior
near the onset of loss of rigidity \cite{gn0,gn1,gn2,gn3,gn4}: The static pressure, shear modulus, and excess coordination number vanish at the onset,
while the first peak of the radial distribution function diverges near the rigidity transition \cite{gr0,gr1,gr2,gr3}
(such a divergence is specific to static packings, which is smoothed out once temperature is imposed to the system \cite{th0,th1,th2}).
In addition, normal mode analyses have revealed excessive low-frequency modes or soft modes near the transition,
implying large-scale collective motions of constituent particles without any changes of elastic energy \cite{vm0,vm1,vm2,vm3}.
Then, introducing the onset as a \emph{jamming packing fraction},\ $\phi_J$,\ a wide variety of thermal (e.g.\ glasses)
and athermal (e.g.\ granular materials) soft particles is unified in a phase diagram \cite{ph0,ph1,ph2},
while there are still some discussions about finite size effects \cite{finite0,finite1} and the uncertainty of $\phi_J$ \cite{nishant}.

%
%
Mechanical properties of soft particle packings have their microscopic origin in complex networks of interparticle (or contact) forces,\ i.e.\ \emph{force-chain networks} \cite{ps0}.
Therefore, any macroscopic quantity can be deduced from statistical averages over the \emph{probability distribution functions} (PDFs) of contact forces,
which have been widely investigated through experiments \cite{ps0,ps1,ps2,ps3,ps4,ps5,ps6,ps7,ps8} and molecular dynamics (MD) simulations of frictionless
\cite{pdf_frl0,pdf_frl1,pdf_frl2,pdf_frl3,pdf_frl4} or frictional particles \cite{pdf_frc0,pdf_frc1,pdf_frc2,pdf_frc3}.
In general, the PDFs are asymmetric and cannot be described by conventional distribution functions \cite{gu4}
such that there is still much debate about their tails \cite{gu5} and asymptotic behavior near the zero contact force \cite{gu3,wyart0,wyart1,wyart2}.
As a result, there have been many attempts to establish a statistical mechanics of ``static" particle packings.
Here, the main idea is that static configurations for the same packing fraction are assumed to be equiprobable,
where the density of states or PDF of forces is to be calculated from appropriate ensembles satisfying some mechanical constraints,\ e.g.\
Edwards' entropy \cite{ed0,ed1,ed2,ed3,ed4,ed5,ed6,ed7}, entropy maximization \cite{em0,em1,em2,em3,em4,em5,ft0},
force ensembles \cite{en3,en4,en5,en6,en7,en8}, stress ensembles \cite{ft2,ft3,ft4}, or others \cite{en0,en1,ft1}.

%
%
When global deformations or \emph{affine deformations} \cite{emt0} are applied to soft particle packings,
however, the particles rearrange to rest on other stable (and more favorable) configurations.
Such rearrangements,\ i.e.\ \emph{non-affine displacements}, of constituent particles, cause anomalous mechanical responses of soft particle packings,
especially, \emph{irreversible responses} to quasi-static deformations \cite{lemaitre,nishant,irrev0,irrev1,saitoh2}.
The degree of non-affinity tends to be strong near the jamming transition \cite{rs0,rs1,rs2}
and vortex-like structures of non-affine displacements have been extensively investigated by experiments \cite{naff_ex0,naff_ex1,naff_ex2}
and MD simulations \cite{naff0,naff1,naff2,naff3,naff4,naff5} with the focus on anisotropy \cite{naff_an0,naff_an1,naff_an2}.
As expected, spatial distributions of large non-affine displacements can be mapped onto localized structures of low-frequency modes \cite{lz0,lz1,lz2,lz3},
which induces notable anomalies of local elastic constants of soft particle packings \cite{local_e0,local_e1,local_e2}.

During such non-affine deformations, we observe the complicated \emph{restructuring} of force-chain networks involving recombinations of contacts,
where existing contacts are broken and new contacts are generated,\ i.e.\ \emph{opening and closing contacts}, respectively.
Then, non-affine responses of contact forces can be regarded as \emph{stochastic} (rather than deterministic affine responses,\ e.g.\ effective medium theory \cite{emt0}).
Therefore, it is natural to describe the evolution of the PDFs by a stochastic model as a consequence of stochastic changes of force-chain networks during non-affine deformations.
Recently, we have proposed a \emph{master equation} for the PDFs \cite{saitoh2},
where the master equation fully describes the non-affine evolution of the PDFs and anomalous responses of macroscopic quantities during deformations.

In this paper, we explain full details of the master equation and investigate further applications with the focus on irreversible responses of soft particle packings.
In the following, we introduce our numerical model in Sec. \ref{sec:ns} and show our results in Sec. \ref{sec:re}.
Then, we discuss and conclude our results in Sec. \ref{sec:discon} and give some technical details in Appendices \ref{app:pdf} and \ref{app:derive}.
%
%
%
\section{Method}
\label{sec:ns}
We use molecular dynamics (MD) simulations of two-dimensional $50:50$ binary mixtures of frictionless soft particles
with the same mass, $m$, and two kinds of radii, $R_i$ and $R_j$ ($R_i/R_j=1.4$).
The normal force between the particles in contact is given by $f_{ij}=kx_{ij}+\eta\dot{x}_{ij}$ with a spring constant, $k$, and viscosity coefficient, $\eta$.
Here, $x_{ij}$ is an overlap between the particles defined as
\begin{equation}
x_{ij} = R_i+R_j-d_{ij}
\label{eq:xij}
\end{equation}
with an interparticle distance, $d_{ij}$, and $\dot{x}_{ij}$ is the relative speed in the normal direction.
A global damping force, $\mathbf{f}_i^\mathrm{damp}=-\eta\mathbf{v}_i$, proportional to the particle's velocity, $\mathbf{v}_i$,
is also introduced to enhance the relaxation, where the damping coefficient is the same with the viscosity coefficient between the particles in contact.

To make static packings of $N$ particles, we randomly distribute them in a $L\times L$ square periodic box, where no particle touches others.
Then, we rescale every radius as
\begin{equation}
R_i(t+\delta t)=\left[1+\frac{\bar{x}-x_\mathrm{m}(t)}{l}\right]R_i(t)
\label{eq:rescaling}
\end{equation}
$(i=1,\dots,N)$, where $t$, $\delta t$, $\bar{x}$, and $x_\mathrm{m}(t)$ are time, an increment of time,
a target value of averaged overlap, and the averaged overlap at time $t$, respectively.
Here, we use a long length scale $l=10^2\bar{\sigma}$ with the mean diameter, $\bar{\sigma}$, to rescale each radius gently.
We confirmed that static packings prepared with longer length scales, $l=10^3\bar{\sigma}$ and $10^4\bar{\sigma}$, give the same results
concerning critical scaling near jamming \cite{gn3}, while we cannot obtain the same results with a shorter length scale, $l=10\bar{\sigma}$.
During the rescaling, each radius increases (decreases) if the averaged overlap is smaller (larger) than the target value,
$x_\mathrm{m}(t)<\bar{x}$ ($x_\mathrm{m}(t)>\bar{x}$), so that the averaged overlap will finally converge to $\bar{x}$.
Note that the ratio between different radii does not change by the rescaling,\ i.e. $R_i(t+\delta t)/R_j(t+\delta t)=R_i(t)/R_j(t)$.
We then stop the rescaling when every acceleration of particles drops below a threshold, $10^{-6}k\bar{\sigma}/m$, and assume that the system is static.

We apply an isotropic compression or decompression to the prepared packings by rescaling every radius as
\begin{equation}
R'_i = \sqrt{1\pm\frac{\delta\phi}{\phi}}R_i
\label{eq:R'i}
\end{equation}
so that the area fraction increases or decreases from $\phi$ to $\phi\pm\delta\phi$.
Then, we relax the system until every acceleration of particles drops below the threshold again.

In our simulations, distances from jamming are determined by the linear scaling of averaged overlap, $\bar{x}(\phi)=A(\phi-\phi_J)$ \cite{gn3},
where we estimate the jamming density as $\phi_J=0.8458\pm10^{-4}$ with the critical amplitude, $A=(0.9\pm0.003)\bar{\sigma}$, from our $50$-sample simulations of $N=8192$ particles.
We also prepared $20$ samples for smaller systems ($N=512,2048$) and $2$ samples for the largest one ($N=32768$) by changing the initial configurations,
while we mainly report the results of $N=8192$ since no result depends on the system size (see Fig.\ \ref{fig:com_pow_ssize}).
%
%
%
\section{Results}\label{sec:re}
In this section, we introduce a master equation for the PDFs of particle overlaps.
The master equation describes microscopic changes of force-chain networks during quasi-static deformations
and can be connected to the constitutive relations through the derivative of the PDFs.
We first study microscopic responses of force-chain networks to quasi-static isotropic (de)compressions (Sec.\ \ref{sub:micro})
and then introduce a master equation for the PDFs (Sec.\ \ref{sub:master}).
To formulate the master equation, we quantify mean values and fluctuations of overlaps (Sec.\ \ref{sub:mean})
and numerically determine transition rates in the master equation (Sec.\ \ref{sub:cpd}).
We validate our framework by comparing numerical solutions of the master equation with MD simulations (Sec.\ \ref{sub:numeric}),
where irreversible responses of soft particle packings to cyclic loading are also examined (Sec.\ \ref{sub:macro}).
\subsection{Microscopic responses}
\label{sub:micro}
At microscopic scales in soft particle packings, mechanical responses to quasi-static deformations are probed as restructuring of force-chain networks,
where complicated particle rearrangements cause the recombination of force-chains,\ i.e.\ opening and closing contacts.
To take into account such opening and closing contacts, we employ the Delaunay triangulation (DT) of particle packings as shown in Fig.\ \ref{fig:delaunay},
where not only the particles in contacts, but also the nearest neighbors without contacts,\ i.e.\ the particles in \emph{virtual contacts}, are connected by the Delaunay edges.
We then generalize overlaps,\ Eq.\ (\ref{eq:xij}), as
\begin{equation}
x_{ij}\equiv R_i+R_j-D_{ij}
\label{eq:go}
\end{equation}
with the Delaunay edge length, $D_{ij}$, where the overlaps (or gaps) between particles in virtual contacts ($R_i+R_j<D_{ij}$) are defined as negative values.
Because the DT is unique for each packing, virtual contacts are uniquely determined.
%
\begin{figure}
\caption{(Color online)
The Delaunay triangulation (DT) of a soft particle packing, where the red and blue solid lines connect the particles in contacts and virtual contacts, respectively.
The width of red solid lines is proportional to the strength of the interparticle force, where the number of particles is $N=512$.
\label{fig:delaunay}}
\end{figure}

If we apply isotropic compression or decompression to the system by Eq.\ (\ref{eq:R'i}), every generalized overlaps (not only contacts, but also virtual contacts) changes to
\begin{equation}
x_{ij}^\mathrm{affine} \simeq x_{ij}\pm\frac{D_{ij}}{2\phi}\delta\phi~,
\label{eq:xij_affine}
\end{equation}
where we neglected the higher order terms proportional to $x_{ij}\delta\phi$ and $\delta\phi^2$
\footnote{An overlap after affine deformation is given by
$x_{ij}^\mathrm{affine}=R'_i+R'_j-d_{ij}=\sqrt{1\pm\delta\phi/\phi}(R_i+R_j)-d_{ij}\simeq R_i+R_j-d_{ij}\pm(R_i+R_j)\delta\phi/2\phi\simeq x_{ij}\pm d_{ij}\delta\phi/2\phi$,
where we used $R_i+R_j=d_{ij}+x_{ij}$ and neglected the higher order terms proportional to $x_{ij}\delta\phi$ and $\delta\phi^2$.}.
However, particles are randomly arranged and each force balance is broken by the affine deformation.
Then, the particles move and the system relaxes to a new equilibrium state, where non-affine displacements during the relaxation
(Fig.\ \ref{fig:displacement}) cause complicated changes of contacts, including opening and closing contacts
\footnote{Note that our systems do not undergo \emph{structural relaxations} after (de)compression, where most particles do not jump out of cages.
We also checked that the response to compression does not depend on the protocols,\ e.g.\ an overdamped dynamics.}.
%
\begin{figure}
\caption{(Color online)
Non-affine displacements of $N=8192$ particles (the red arrows),
where the color coordinate (from 0 to 1) represents the magnitude of non-affine displacements scaled by the maximum value.
\label{fig:displacement}}
\end{figure}

\begin{figure}
\includegraphics[width=\columnwidth]{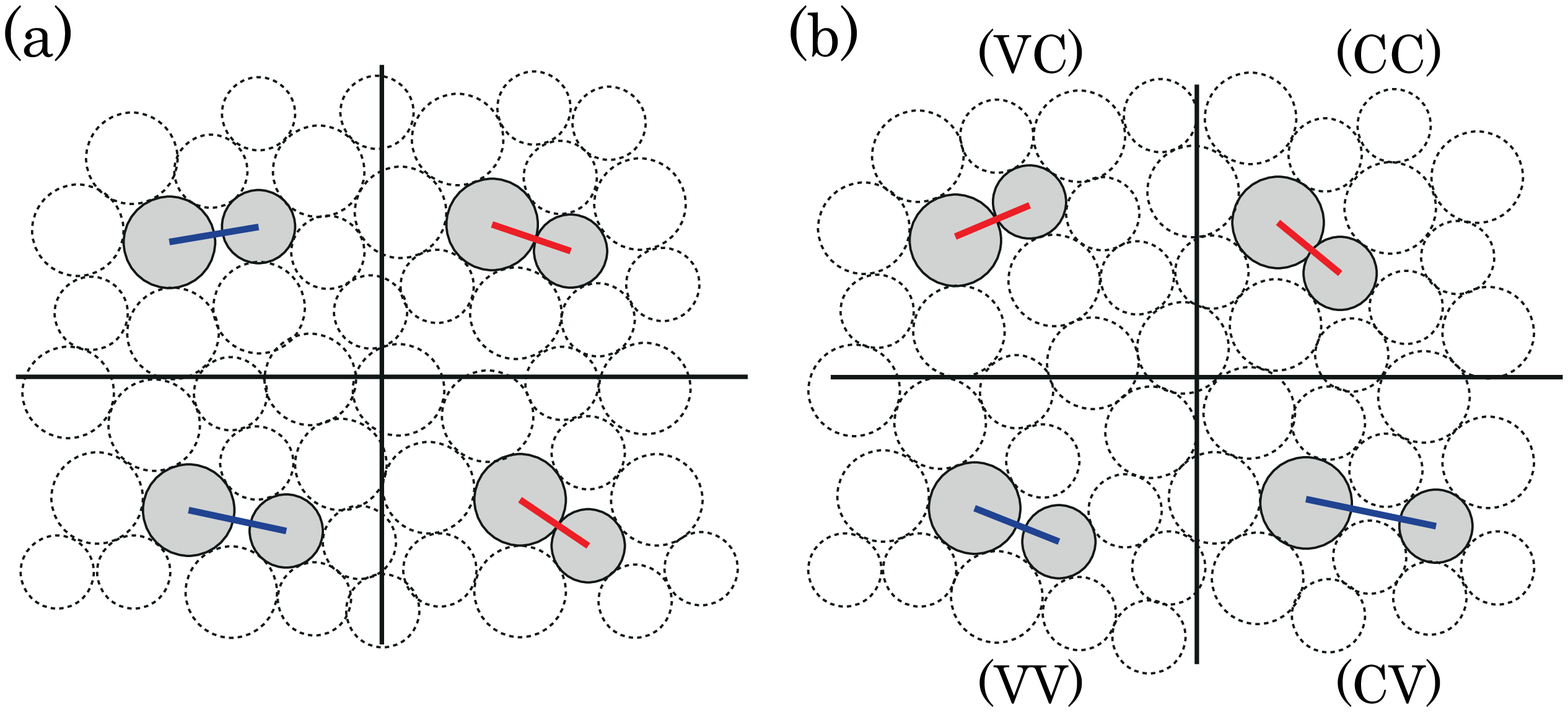}
\caption{(Color online)
Sketches of contacts (a) before and (b) after deformation, where four kinds of transitions are displayed:
(CC) contact-to-contact, (VV) virtual-to-virtual, (CV) contact-to-virtual, and (VC) virtual-to-contact, respectively.
\label{fig:contacts}}
\end{figure}

After the relaxation, overlaps change to new values, $x'_{ij}\neq x_{ij}^\mathrm{affine}$, that is non-affine responses of overlaps.
As shown in Fig.\ \ref{fig:contacts}, there are only four kinds of changes from $x_{ij}$ to $x'_{ij}$:
A positive overlap, $x_{ij}>0$, remains positive, $x'_{ij}>0$, or a negative overlap, $x_{ij}<0$, stays negative, $x'_{ij}<0$, such that contacts are neither generated nor broken.
We call these changes ``\emph{contact-to-contact} (CC)" and ``\emph{virtual-to-virtual} (VV)" transitions, respectively.
On the other hand, if a positive overlap changes to a negative one and a negative overlap becomes positive, an existing contact is broken and a new contact is generated, respectively.
We name these changes ``\emph{contact-to-virtual} (CV)" and ``\emph{virtual-to-contact} (VC)" transitions, respectively.
%

In the following, we scale the generalized overlaps by the averaged overlap before deformation, $\bar{x}(\phi)$.
Then, the affine response, Eq.\ (\ref{eq:xij_affine}), is scaled as
\begin{equation}
\xi^\mathrm{affine}=\xi\pm B_a\gamma
\label{eq:xi_affine}
\end{equation}
which is a linear function of a scaled overlap before deformation, $\xi\equiv x_{ij}/\bar{x}(\phi)$.
Here, we omit the subscript, $ij$, after the scaling.
On the right-hand-side of Eq.\ (\ref{eq:xi_affine}), the offset is proportional to a \emph{scaled strain increment},
\begin{equation}
\gamma\equiv\frac{\delta\phi}{\phi-\phi_J}~,
\label{eq:gamma}
\end{equation}
with the amplitude defined as $B_a\equiv D_{ij}/(2A\phi)$.
Similarly, the non-affine response is scaled as $\xi'\equiv x'_{ij}/\bar{x}(\phi)\neq\xi^\mathrm{affine}$.

\begin{figure}
\includegraphics[width=8cm]{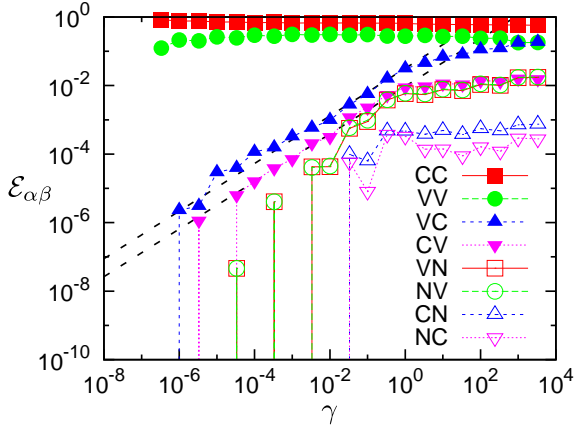}
\caption{(Color online)
Probabilities of contact changes, $\mathcal{E}_{\alpha\beta}$, plotted against the scaled strain increment, $\gamma$,
where we take $50$-sample averages of $N=8192$ particles for each value of $\gamma$,\ i.e.\ for each combination of $\delta\phi$ and $\phi-\phi_J$.
The subscripts, $\alpha$ and $\beta$ ($=C,V,N$), represent the status before ($\alpha$) and after ($\beta$) compressions,
where $C$, $V$, and $N$ mean that the particles are in contact, in virtual contact, and in neither contact nor virtual contact, respectively.
Different symbols represent different contact changes as listed in the legend.
The two dashed lines are power law fits for the probabilities, $\mathcal{E}_{VC}\simeq3.17\times10^{-2}\gamma^{0.7}$ and $\mathcal{E}_{CV}\simeq0.94\times10^{-2}\gamma^{0.7}$.
\label{fig:delsum}}
\end{figure}

Note that we only analyze contact changes occurring on the generalized force-chain networks.
If the applied strain increment is too large, we also observe that particles, which were neither in contact nor in virtual contact, are connected by Delaunay edges after deformation, and vice versa.
However, we confirm that such rare events are below our statistical significance for the whole range of scaled strain increments used in our MD simulations
\footnote{Thus, we will not consider any source- and sink-terms in the master equation in Sec.\ \ref{sub:master}.}.
As shown in Fig.\ \ref{fig:delsum}, the probability of finding virtual contacts, which are broken to or generated from the pairs neither in contacts nor in virtual contacts,\
i.e.\ $\mathcal{E}_{VN}$ or $\mathcal{E}_{NV}$, is less than $3\%$ (the open squares and circles).
Similarly, the probability of finding contacts which are broken to or generated from such pairs,\ i.e.\ $\mathcal{E}_{CN}$ or $\mathcal{E}_{NC}$, is less than $0.1\%$ (the open triangles).
Though the two cases, (CC) and (VV), dominate the number of contact changes, the number of closing and opening contacts,
(VC) and (CV), asymptotically decays to zero with the scaled strain increment as $\mathcal{E}_{VC},\mathcal{E}_{CV}\sim\gamma^{0.7}$ if $\gamma\le1$.
Therefore, the system exhibits non-affine deformations even if $\gamma$ is very small.
%
%
\subsection{A master equation}
\label{sub:master}
\begin{figure}
\includegraphics[width=8cm]{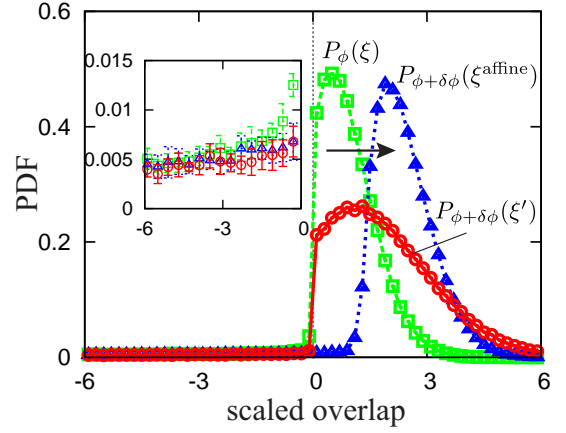}
\caption{(Color online)
The PDFs of scaled overlaps, $P_\phi(\xi)$ (squares), $P_{\phi+\delta\phi}(\xi_\mathrm{affine})$ (triangles), and $P_{\phi+\delta\phi}(\xi')$ (circles),
obtained from our MD simulations of $N=8192$ particles with the distance from jamming, $\phi-\phi_J=1.2\times10^{-3}$.
Here, we apply an isotropic compression with $\delta\phi=4\times10^{-4}$,\ i.e.\ $\gamma=\delta\phi/\left(\phi-\phi_J\right)=0.33$.
The inset is the zoom-in to the PDFs of virtual contacts, where the error-bars are obtained from $50$-sample averages.
Because of the bidispersed diameters of particles, the discontinuous gap around zero in the initial PDF, $P_\phi(\xi)$,
is smoothed out after the affine deformation, $P_{\phi+\delta\phi}(\xi_\mathrm{affine})$.
\label{fig:pdfs}}
\end{figure}

The restructuring of force-chain networks attributed to the contact changes, (CC), (VV), (CV), and (VC),
is well captured by the PDFs of scaled overlaps, $P_\phi(\xi)$, where the subscript represents the area fraction.
Because the total number of contacts and virtual contacts is conserved during deformations, the PDFs are normalized as
\begin{equation}
\int_{-\infty}^\infty P_\phi(\xi)d\xi = 1~.
\label{eq:norm_pdf}
\end{equation}
Figure \ref{fig:pdfs} shows the PDFs obtained from our MD simulations before compression, $P_\phi(\xi)$, after affine deformation, $P_{\phi+\delta\phi}(\xi^\mathrm{affine})$,
and after non-affine deformation, $P_{\phi+\delta\phi}(\xi')$, where the affine compression just shifts the initial PDF to the positive direction,
while the non-affine deformation broadens it in positive overlaps and maintain a discontinuous ``jump" around zero \cite{th1}.
Note that, however, the PDF after non-affine deformation in negative overlaps is comparable to that after affine deformation (see the inset in Fig.\ \ref{fig:pdfs}).
%

To describe such the non-affine evolution of the PDFs, we assume that transitions between overlaps (from $\xi$ to $\xi'$) are \emph{Markov processes}
such that we can connect the PDF after non-affine deformation to that before compression through the \emph{Chapman-Kolmogorov equation} \cite{vanKampen},
\begin{equation}
P_{\phi+\delta\phi}(\xi') = \int_{-\infty}^\infty W(\xi'|\xi)P_\phi(\xi)d\xi~,
\label{eq:chapman-kolmogorov}
\end{equation}
where $W(\xi'|\xi)$ is the \emph{conditional probability distribution} (CPD) of the overlaps, $\xi'$, which were $\xi$ before the compression.
In the Chapman-Kolmogorov equation (\ref{eq:chapman-kolmogorov}), the source- and sink-terms caused by the rare events,
$\mathcal{E}_{VN}$, $\mathcal{E}_{NV}$, $\mathcal{E}_{CN}$, and $\mathcal{E}_{NC}$, are not taken into account.
By definition, the CPD is normalized as \cite{vanKampen}
\begin{equation}
\int_{-\infty}^\infty W(\xi'|\xi)d\xi'=1~.
\label{eq:norm_cpd}
\end{equation}
From Eqs.\ (\ref{eq:chapman-kolmogorov}) and (\ref{eq:norm_cpd}), a \emph{master equation} for the PDFs is readily found to be \cite{vanKampen}
\begin{equation}
\frac{\partial}{\partial\phi}P_\phi(\xi') = \int_{-\infty}^\infty\left\{T(\xi'|\xi)P_\phi(\xi)-T(\xi|\xi')P_\phi(\xi')\right\}d\xi~,
\label{eq:master}
\end{equation}
if we introduce a \emph{transition rate} \cite{vanKampen} as
\begin{equation}
T(\xi'|\xi)=\lim_{\delta\phi\rightarrow0}\frac{W(\xi'|\xi)}{\delta\phi}~.
\label{eq:trate}
\end{equation}
The first and second terms in the integral on the right-hand-side of the master equation (\ref{eq:master}) represent the \emph{gain} and \emph{loss} of overlaps, $\xi'$, respectively.
Therefore, the transition rates or CPDs fully determine the statistics of microscopic changes of force-chain networks.

If we multiply Eq.\ (\ref{eq:master}) by $\phi-\phi_J$ and introduce an infinitesimal scaled strain increment as
$\delta\gamma\equiv\delta\phi/(\phi-\phi_J)\ll1$, we obtain an alternative form of the master equation as
\begin{equation}
\frac{\partial}{\partial\gamma}P_\phi(\xi')=\int_{-\infty}^\infty\left[T_\gamma(\xi'|\xi)P_\phi(\xi)-T_\gamma(\xi|\xi')P_\phi(\xi')\right]d\xi~,
\label{eq:master_gamma}
\end{equation}
where the transition rates are now defined as $T_\gamma(\xi'|\xi)=\lim_{\delta\gamma\rightarrow0}W(\xi'|\xi)/\delta\gamma$.

In addition, the master equation for unscaled overlaps, $x=\bar{x}(\phi)\xi$ and $x'=\bar{x}(\phi)\xi'$, can be written as
\begin{equation}
\frac{\partial}{\partial\phi}P^\ast_\phi(x')=\int_{-\infty}^\infty\left[T^\ast(x'|x)P^\ast_\phi(x)-T^\ast(x|x')P^\ast_\phi(x')\right]dx~,
\label{eq:non_master}
\end{equation}
where the unscaled PDF, unscaled transition rate, and unscaled increment are given by $P^\ast_\phi(x)=\bar{x}(\phi)^{-1}P_\phi(\xi)$,
$T^\ast(x'|x)=\bar{x}(\phi)^{-1}T(\xi'|\xi)$, $dx=\bar{x}(\phi)d\xi$, respectively.

The PDFs can be connected to macroscopic quantities through the \emph{$n$-th moment} of positive overlaps,
\begin{equation}
\mu_n\equiv\int_0^\infty x^n P_\phi(x)dx=\int_0^\infty x^n P_\phi(\xi)d\xi~,
\label{eq:moment}
\end{equation}
where we used the relation, $P_\phi(x)dx=P_\phi(\xi)d\xi$.
Then, the coordination number, $z$, averaged overlap, $\bar{x}$, and static pressure, $p$, are given by the first three moments as
\begin{eqnarray}
z &=& \frac{2N_E}{N}\mu_0~,\label{eq:momt_z}\\
\bar{x} &=& \mu_1~,\label{eq:momt_x}\\
p &=& \frac{kN_E}{L^2}\mu_0\left(\bar{\sigma}\mu_1-\mu_2\right)~,\label{eq:momt_p}
\end{eqnarray}
respectively, where $N_E$ is the total number of the Delaunay edges
\footnote{The number of contacts and virtual contacts are given by $N_C=M_0N_E$ and $N_V=(1-M_0)N_E$, respectively,
so that the coordination number is given by $z=2N_C/N=2M_0N_E/N$.
The mean overlap is equivalent to the first moment, $\bar{x}(\phi)=M_1$, and then the static pressure is given by
$p=(k/L^2)<x_{ij}d_{ij}>=(k/L^2)<(R_i+R_j)x_{ij}-x_{ij}^2>\simeq(kN_C/L^2)(\bar{\sigma}<x>-<x^2>)=(kN_E/L^2)M_0(\bar{\sigma}M_1-M_2)$,
where we neglected the weak correlation between the sum of radii, $R_i+R_j$, and overlap, $x_{ij}$.
}.
%
\subsection{Mean and fluctuations}
\label{sub:mean}
\begin{figure}
\includegraphics[width=\columnwidth]{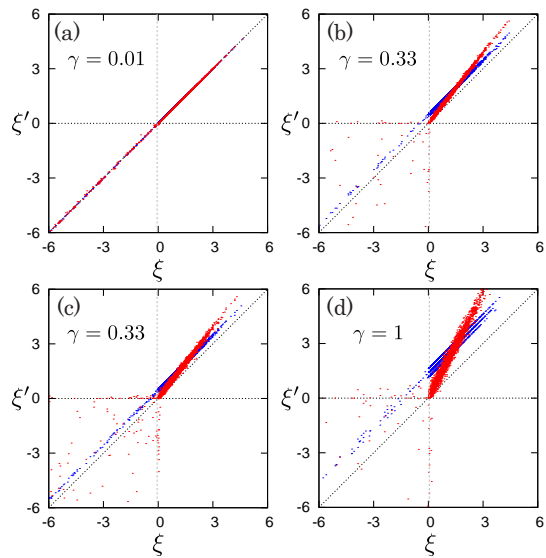}
\caption{(Color online) Scatter plots of scaled overlaps under \emph{compression},
where the blue and red dots are $\xi^\mathrm{affine}$ and $\xi'$ plotted against $\xi$, respectively.
Distances from jamming are $\phi-\phi_J=$ (a) $4\times10^{-3}$, (b) $1.2\times10^{-4}$, (c) $1.2\times10^{-3}$, and (d) $4\times10^{-4}$, respectively,
while applied strain increments are $\delta\phi=4\times10^{-5}$ ((a) and (b)) and $4\times10^{-4}$ ((c) and (d)), respectively,
such that scaled strain increments are given by $\gamma=$ (a) $0.01$, (b) $0.33$, (c) $0.33$, and (d) $1$, respectively.
Here, the number of particles is $N=8192$ ($1$ sample).
\label{fig:emsc_com}}
\end{figure}

%
Because the transition rates or CPDs are defined as distributions of $\xi'$ around their mean values,
we first measure the mean and fluctuations of $\xi'$ through \emph{scatter plots} of scaled overlaps.
Figure \ref{fig:emsc_com} displays the scatter plots obtained from our MD simulations under compressions.
In this figure, the four kinds of transitions displayed in Fig.\ \ref{fig:contacts} are mapped onto four regions:
(CC) $\xi,\xi'>0$, (VV) $\xi,\xi'<0$, (CV) $\xi>0$, $\xi'<0$, and (VC) $\xi<0$, $\xi'>0$, respectively.
Though scaled overlaps after affine deformation, $\xi^\mathrm{affine}$, are described by the deterministic equation (\ref{eq:xi_affine}),
those after non-affine deformation distribute around their mean values with finite fluctuations,\ i.e.\ non-affine responses of overlaps are \emph{stochastic}.
The differences between affine and non-affine responses are always present, but not visible
if the applied strain is small or the system is far from jamming,\ i.e.\ if $\gamma\ll1$ (Fig.\ \ref{fig:emsc_com}(a)),
while $\xi'$ deviates more from $\xi^\mathrm{affine}$ and data points are more dispersed if $\gamma\gg1$ (Fig.\ \ref{fig:emsc_com}(d)).
%
\begin{figure}
\includegraphics[width=\columnwidth]{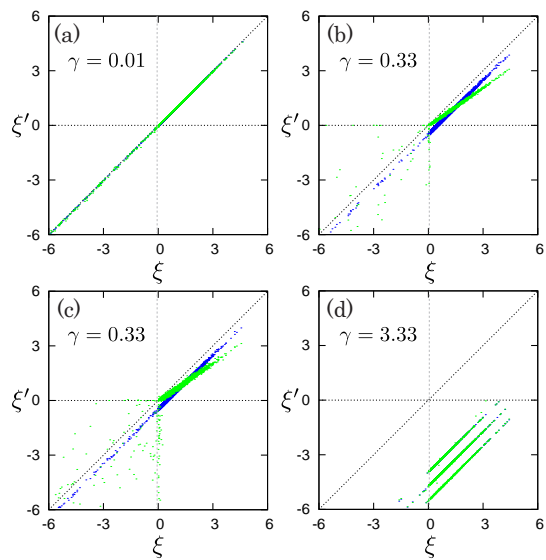}
\caption{(Color online) Scatter plots of scaled overlaps under \emph{decompression},
where the blue and green dots are $\xi^\mathrm{affine}$ and $\xi'$ plotted against $\xi$, respectively.
Applied strain increments are $\delta\phi=4\times10^{-5}$ ((a) and (b)) and $4\times10^{-4}$ ((c) and (d)), respectively,
where distances from jamming, magnitudes of scaled strain increments, and the number of particles are as in Fig.\ \ref{fig:emsc_com}.
(d) If $\gamma>1$, an unjamming transition happens.
\label{fig:emsc_dec}}
\end{figure}

\begin{figure}
\includegraphics[width=6cm]{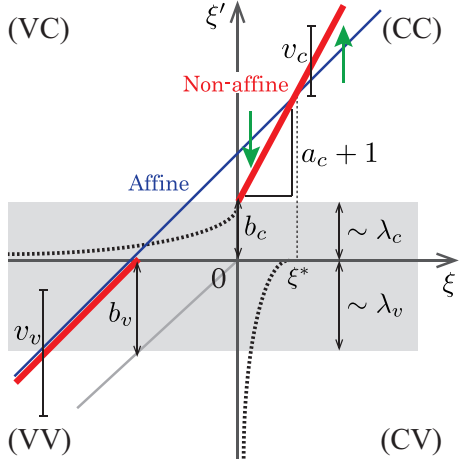}
\caption{(Color online)
A schematic picture of the difference between affine and non-affine responses of scaled overlaps,
where the blue and red solid lines represent $\xi^\mathrm{affine}$ (for small and large particles) and linear functions, $m_l(\xi)$, respectively.
In (CC), $\xi^\mathrm{affine}$ and $m_l(\xi)$ intersect at $\xi=\xi^\ast$, where the green arrows represent the decrease and increase of scaled overlaps during relaxation.
The excess slope in (CC), $a_c$, and all the dimensionless lengths, $b_l$, $v_l$, and $\lambda_l$, are proportional to $\gamma$,
where $\lambda_c$ and $\lambda_v$ represent typical \emph{penetration lengths} of new contacts and new virtual contacts, respectively.
\label{fig:scat}}
\end{figure}

In (CC) and (VV) regions, the mean values of $\xi'$ can be fitted by \emph{linear functions} of $\xi$,
\begin{equation}
m_l(\xi) = (a_l+1)\xi+b_l~,
\label{eq:mzt}
\end{equation}
where the subscripts, $l=c$ and $v$, represent the mean values in (CC) and (VV), respectively.
We also introduce standard deviations of $\xi'$ from their mean values as $v_l$, which are found to be almost independent of $\xi$.
Then, the systematic deviation from the affine response, $\xi^\mathrm{affine}$, is quantified by the coefficients, $a_l$, $b_l$, and $v_l$ as summarized in Fig. \ref{fig:scat}.
Note that the affine response, Eq.\ (\ref{eq:xi_affine}), is recovered if $a_l=v_l=0$ and $b_l=B_a\gamma$.

\begin{figure}
\includegraphics[width=\columnwidth]{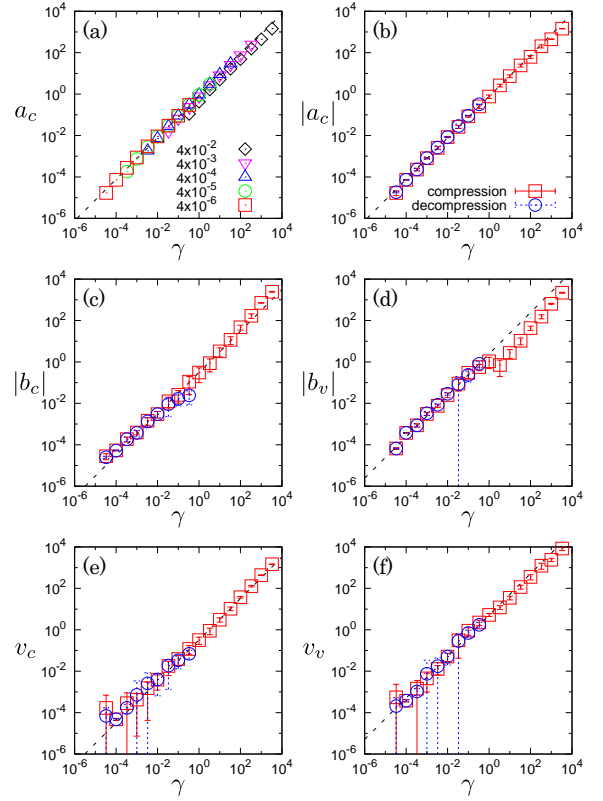}
\caption{(Color online)
(a) Data collapse of the excess slope in (CC), $a_c$, plotted against $\gamma=\delta\phi/(\phi-\phi_J)$ under compression,
where different symbols represent $\delta\phi$ as listed in the legend.
(b)-(f): Double logarithmic plots of the coefficients, (b) $|a_c|$, (c) $|b_c|$, and (d) $|b_v|$,
and standard deviations, (e) $v_c$ and (f) $v_v$, under compression (squares) and decompression (circles),
where we averaged the data over different combinations of $\delta\phi$ and $\phi-\phi_J$ for the same $\gamma$.
Here, the dashed lines represent the linear scaling against $\gamma$,\ i.e.\ Eq.\ (\ref{eq:scaling})
(the vertical dotted lines represent error-bars comparable to the mean values).
\label{fig:com_dec_pow}}
\end{figure}

\begin{figure}
\includegraphics[width=\columnwidth]{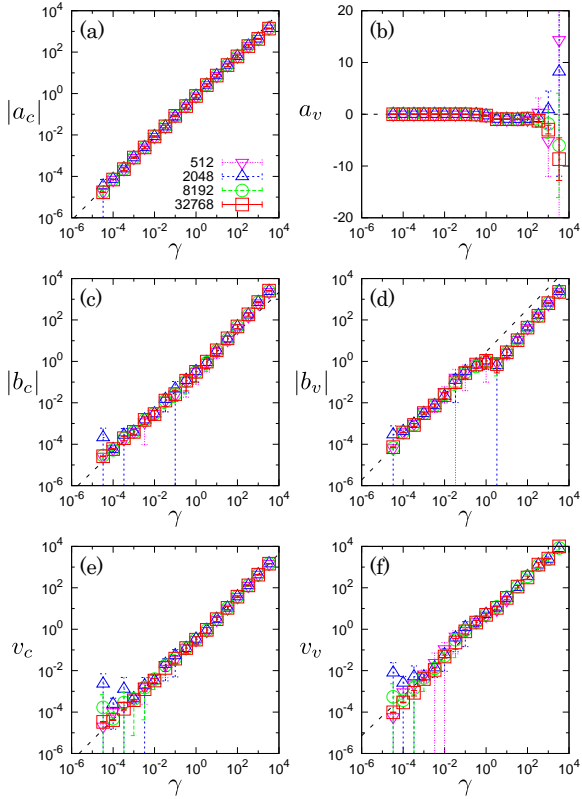}
\caption{(Color online)
The system size dependence of the coefficients under compression, (a) $|a_c|$, (b) $a_v$, (c) $|b_c|$, (d) $|b_v|$, (e) $v_c$, and (f) $v_v$,
where the dotted lines represent the linear scalings of $\gamma$,\ i.e.\ Eq.\ (\ref{eq:scaling}), except for $a_v\simeq0$.
The different symbols represent different system sizes, $N$, as listed in the legend in (a).
\label{fig:com_pow_ssize}}
\end{figure}

As we observed in the scatter plots (Fig.\ \ref{fig:emsc_com}), the difference between affine and non-affine responses increases with the scaled strain increment, $\gamma$.
From our simulations, we find that all the coefficients, except for $a_v\simeq0$, \emph{linearly} increase with $\gamma$,
where all data with a wide variety of $\delta\phi$ and $\phi-\phi_J$ collapse onto a linear scaling of $\gamma$ (Fig.\ \ref{fig:com_dec_pow}(a)).
As shown in Figs.\ \ref{fig:com_dec_pow}(b)-(f), all the coefficients (the open squares) are well described by linear scalings of the scaled strain increment,
\begin{equation}
a_l = A_l\gamma~,\hspace{3mm} b_l = B_l\gamma~,\hspace{3mm} v_l = V_l|\gamma|~,
\label{eq:scaling}
\end{equation}
where the scaling amplitudes, $A_l$, $B_l$, and $V_l$, estimated in the fitting range, $10^{-6}\le\gamma\le1$, are listed in Table \ref{tab:scaling}.
Because $a_v\simeq0$ and $B_v\approx B_a$($\simeq1.3$ for small and large particles),
virtual contacts almost behave affine in average except for their huge fluctuations ($V_v\gg V_c$).
In contrast, $B_c$ is always smaller than $B_a$ such that $m_c(\xi)$ intersects $\xi^\mathrm{affine}$
at a characteristic scaled overlap, $\xi^\ast=(B_a-B_c)/A_c\simeq1.4$ (which is independent of $\gamma$).
This leads to small responses ($\xi'<\xi_\mathrm{affine}$) of small overlaps ($\xi<\xi^\ast$) and vice versa as indicated by the green arrows in Fig.\ \ref{fig:scat},
implying preferred tangential and hindered normal displacements as a sign of non-affine deformations \cite{rs1}.
As shown in Fig.\ \ref{fig:com_pow_ssize}, the linear scalings, Eq.\ (\ref{eq:scaling}), do not depend on the system size.
%
%

The linear scalings, Eq.\ (\ref{eq:scaling}), are retained under decompression,\ i.e.\ for $-\gamma$.
Figures \ref{fig:emsc_dec} displays scatter plots of scaled overlaps under decompression, where an \emph{unjamming transition} happens if $\gamma>1$ (Fig.\ \ref{fig:emsc_dec}(d))
\footnote{The unjamming transition is beyond the reach of our method, because the source- and sink-terms cannot be neglected in the master equation and $\gamma>1$ is not small enough.}.
Here, we also fit the mean values of $\xi'$ by $m_l(\xi)$ and quantify the fluctuations by standard deviations, $v_l$.
Then, we find that the excess slope and offsets are negative, $a_c, b_l<0$, while $a_v\simeq0$ and $v_l>0$ as in the case of compression.
As shown in Figs.\ \ref{fig:com_dec_pow}(b)-(f), their absolute values (the open circles) are well described by the same linear scalings for compression
so that Eq.\ (\ref{eq:scaling}) can be used for both compression and decompression
(because $a_c$ and $b_l$ are linear against $\gamma$,\ i.e.\ $a_c,b_l\sim\gamma$, they are positive and negative under compression, $\gamma$, and decompression, $-\gamma$, respectively).

In contrast to (CC) and (VV), the data of $\xi'$ in (VC) and (CV) are concentrated in narrow regions
(the inside of the dashed lines in Fig.\ \ref{fig:scat}, left and below, respectively).
Here, $\xi^\mathrm{affine}$ linearly increases with $\xi$ in (VC) and there is no data of $\xi^\mathrm{affine}$ in (CV), because affine compressions do not generate any opening contacts.
Similarly, affine decompressions do not generate any closing contacts.
%
\begin{table*}
\caption{Scaling amplitudes in Eqs.\ (\ref{eq:scaling}), $q$-indices in Eq.\ (\ref{eq:q-gauss}), and dimensionless length scales in Eqs.\ (\ref{eq:cpd4}) and (\ref{eq:cpd3}).
\label{tab:scaling}}
\begin{ruledtabular}
\begin{tabular}{cccccc}
$l$ & $A_l$ & $B_l$ & $V_l$ & $q_l$ & $\lambda_l$ \\ \hline
$c$ & $0.76\pm2.4\times10^{-3}$ & $0.24\pm3.3\times10^{-3}$ & $0.32\pm4.8\times10^{-3}$ & $1.13\pm0.13$ & $6.10\pm0.33$ \\
$v$ & $0.00\pm8.3\times10^{-3}$ & $1.80\pm8.4\times10^{-2}$ & $4.41\pm1.9\times10^{-1}$ & $1.39\pm0.18$ & $0.65\pm0.04$ \\
\end{tabular}
\end{ruledtabular}
\end{table*}
%
\subsection{Conditional probability distributions}
\label{sub:cpd}
The statistics of restructuring of force-chain networks is fully described by the transition rates
in the master equation (\ref{eq:master}) or CPDs in the Chapman-Kolmogorov equation (\ref{eq:chapman-kolmogorov}).
For example, the CPD for affine deformation is given by a delta function as
\begin{equation}
W_\mathrm{affine}(\xi'|\xi)=\delta(\xi'-\xi^\mathrm{affine})~,
\end{equation}
which just shifts the PDF by $B_a\gamma$,\ i.e.\ $P_{\phi+\delta\phi}(\xi)=P_\phi(\xi-B_a\gamma)$, as shown in Fig.\ \ref{fig:pdfs}.
However, the CPDs for non-affine deformations distribute around the mean values, $m_l(\xi)$, with finite widths, $v_l$.
In the following, we determine the CPDs after non-affine deformations from our results of MD simulations,
where we explain how to calculate the CPDs from numerical data in Appendix \ref{app:pdf}.

First, we determine the CPDs in (CC) and (VV) under compression.
Figures \ref{fig:all_cpd}(a) and (b) show the CPDs in (CC) and (VV), respectively,
where all data are \emph{symmetric} around the mean values and well collapse if we multiply the CPDs and distances from the mean values,
$\Xi_l\equiv\xi'-m_l(\xi)$, by $\gamma$ and $1/\gamma$, respectively.
In these figures, the CPDs in (CC) and (VV) are well described by the solid lines,
\begin{eqnarray}
\gamma W_{CC}(\xi'|\xi) &=& f_c(\Xi_c/\gamma)~, \label{eq:cpd1}\\
\gamma W_{VV}(\xi'|\xi) &=& f_v(\Xi_v/\gamma)~, \label{eq:cpd2}
\end{eqnarray}
respectively, where
\begin{equation}
f_l(x) = \frac{1}{c(q_l)}\left[1+\frac{x^2}{n(q_l)V_l^2}\right]^{\frac{1}{1-q_l}}
\label{eq:q-gauss}
\end{equation}
is the \emph{q-Gaussian distribution} \cite{q-Gaussian0,q-Gaussian1,q-Gaussian2}.
In Eq.\ (\ref{eq:q-gauss}), the functions are introduced as $n(t)=(t-3)/(1-t)$ and $c(t)=V_l\sqrt{n(t)}B\left(1/2,n(t)/2\right)$ with the beta function, $B(x,y)$.
In Table \ref{tab:scaling}, we list the \emph{q-indices} ($1<q_l<3$) which determine shapes of the CPDs,
where the normal (Gaussian) distribution is recovered in the limit of $q_l\rightarrow1$
\footnote{The $q$-Gaussian distribution is equivalent to \emph{Student's t-distribution},
\[f_l(x) = \frac{1}{c_l}\left[1+\frac{x^2}{n_lV_l^2}\right]^{-\frac{n_l+1}{2}}~,\]
where $c_l=V_l\sqrt{n_l}B\left(1/2,n_l/2\right)$ and its index, $n_l$, is connected with the $q$-index, $q_l$, by $n_l=(q_l-3)/(1-q_l)$.
The cumulative Student's $t$-distribution is given by
\[F_l(x)=\int_{-\infty}^x f_l(x')dx'=1-\frac{1}{2}B\left[\frac{n_l}{x^2+n_l};\frac{n_l}{2},\frac{1}{2}\right]~.\]}.

Next, we determine the CPDs in (CV) and (VC) under compression,
where the normalization condition, Eq.\ (\ref{eq:norm_cpd}), is satisfied in $\xi>0$ and $\xi<0$ as
\begin{eqnarray}
\int_0^\infty W_{CC}(\xi'|\xi)d\xi' + \int_{-\infty}^0 W_{CV}(\xi'|\xi)d\xi' &=& 1~, \label{eq:norm_pos}\\
\int_0^\infty W_{VC}(\xi'|\xi)d\xi' + \int_{-\infty}^0 W_{VV}(\xi'|\xi)d\xi' &=& 1~, \label{eq:norm_neg}
\end{eqnarray}
respectively.
Figures \ref{fig:all_cpd}(c) and (d) show the CPDs in (CV) and (VC), respectively, where all results collapse after the same scaling as for those in (CC) and (VV).
In these figures, the CPDs in (CV) and (VC) are well described by the \emph{exponential distributions},
\begin{eqnarray}
\gamma W_{CV}(\xi'|\xi) &=& F_c\left[\frac{m_c(\xi)}{v_c}\right]\frac{e^{\xi'/\gamma\lambda_v}}{\lambda_v}~, \label{eq:cpd4}\\
\gamma W_{VC}(\xi'|\xi) &=& F_v\left[\frac{m_v(\xi)}{v_v}\right]\frac{e^{-\xi'/\gamma\lambda_c}}{\lambda_c}~,\label{eq:cpd3}
\end{eqnarray}
respectively, where the dimensionless length scales, $\lambda_l$, are listed in Table \ref{tab:scaling}.
On the right-hand-sides of Eqs.\ (\ref{eq:cpd4}) and (\ref{eq:cpd3}), the $\xi$-dependent functions are defined as the incomplete beta function,
\begin{equation}
F_l(x) = \frac{1}{2}B\left[\frac{n(q_l)}{x^2+n(q_l)};\frac{n(q_l)}{2},\frac{1}{2}\right]~,
\label{eq:ibeta}
\end{equation}
which are equivalent to the cumulative distributions of the CPDs in (CC) and (VV),\ i.e.\
\begin{eqnarray}
F_c\left[\frac{m_c(\xi)}{v_c}\right] &=& 1-\int_0^\infty W_{CC}(\xi'|\xi)d\xi'~, \label{eq:Fc}\\
F_v\left[\frac{m_v(\xi)}{v_v}\right] &=& 1-\int_{-\infty}^0 W_{VV}(\xi'|\xi)d\xi'~, \label{eq:Fv}
\end{eqnarray}
respectively.
From Eqs.\ (\ref{eq:Fc}) and (\ref{eq:Fv}), and a relation
\begin{equation}
\int_0^\infty\frac{e^{-\xi'/\gamma\lambda_c}}{\lambda_c}d\xi'=\int_{-\infty}^0\frac{e^{\xi'/\gamma\lambda_v}}{\lambda_v}d\xi'=1~,
\end{equation}
it is confirmed that the distributions defined as Eqs.\ (\ref{eq:cpd1}), (\ref{eq:cpd2}), (\ref{eq:cpd4}), and (\ref{eq:cpd3})
satisfy the normalization conditions of the CPDs, Eqs.\ (\ref{eq:norm_pos}) and (\ref{eq:norm_neg}).
As shown in Figs.\ \ref{fig:all_cpd_xi}(a) and (b), Eqs.\ (\ref{eq:cpd4}) and (\ref{eq:cpd3}) well describe the $\xi$-dependence of the CPDs in (CV) and (VC), respectively.
Because the CPDs in (VC) and (CV) exponentially decay along the $\xi'$-axis,
the dimensionless length scales, $\lambda_c$ and $\lambda_v$ ($\lambda_c\ll\lambda_v$), represent
typical \emph{penetration lengths} for new contacts and \emph{generated gaps} between new virtual contacts.
Therefore, closing and opening contacts contribute to the new PDF, $P_{\phi+\delta\phi}(\xi')$, between $-\lambda_v\lesssim\xi'/\gamma\lesssim\lambda_c$ as shown in Fig.\ \ref{fig:scat}.
Note that, if $\gamma=0$, the CPDs defined as Eqs.\ (\ref{eq:cpd1}), (\ref{eq:cpd2}), (\ref{eq:cpd4}), and (\ref{eq:cpd3})
converge to $W_{CC}=W_{VV}=\delta(\xi-\xi')$ and $W_{CV}=W_{VC}=0$, respectively
\footnote{We used $e^{-1/\gamma}/\gamma\rightarrow0$ for $\gamma\rightarrow0$.},
so that the Chapman-Kolmogorov equation (\ref{eq:chapman-kolmogorov}) does not change the PDF without deformation.

The CPDs under decompression are given by replacing the scaled strain increment, $\gamma$, with $-\gamma$, as in the case of linear scaling, Eq.\ (\ref{eq:scaling}), for decompression.
Figures \ref{fig:all_cpd}(e)-(h) display the CPDs under decompression, where we multiply the CPDs and distances from the mean values by $\gamma$ and $1/\gamma$, respectively.
Figures \ref{fig:all_cpd_xi}(c) and (d) show the $\xi$-dependence of the CPDs in (CV) and (VC) under decompression, respectively.
In these figures, the solid lines correspond to Eqs.\ (\ref{eq:cpd1}), (\ref{eq:cpd2}), (\ref{eq:cpd4}), and (\ref{eq:cpd3}),
such that the CPDs do not change their shapes under compression and decompression.
%
%
\begin{figure*}
\caption{(Color online)
Semi-logarithmic plots of the CPDs in (a) (CC), (b) (VV), (c) (CV), and (d) (VC) under \emph{compression},
and in (e) (CC), (f) (VV), (g) (CV), and (h) (VC) under \emph{decompression}.
Different symbols represent different scaled increments as listed in the legends.
The solid lines are given by the distribution functions, Eqs.\ (\ref{eq:cpd1}), (\ref{eq:cpd2}), (\ref{eq:cpd4}), and (\ref{eq:cpd3}).
The dotted lines in (a) and (e) are Gaussian fits, see Ref.\ \cite{saitoh2}, while those in (b) and (f) are the $q$-Gaussian distributions for (CC).
\label{fig:all_cpd}}
\end{figure*}
\begin{figure}
\caption{(Color online)
Semi-logarithmic plots of the CPDs in (a) (CV) and (b) (VC) under \emph{compression},
and in (c) (CV) and (d) (VC) under \emph{decompression}.
Different symbols represent different scaled increments as given in Fig.\ \ref{fig:all_cpd}.
The solid lines are given by the distribution functions, Eqs.\ (\ref{eq:cpd4}) and (\ref{eq:cpd3}),
and the insets show the corresponding semi-logarithmic plots.
\label{fig:all_cpd_xi}}
\end{figure}
%
%
%
%
%
\subsection{Numerical validations of the master equation}
\label{sub:numeric}
We now unfold the master equation for the PDFs of scaled overlaps, where the transition rates are divided into the four cases,
(CC), (VV), (CV), and (VC), as $T_{CC}(\xi'|\xi)=\lim_{\delta\phi\rightarrow0}W_{CC}(\xi'|\xi)/\delta\phi$, etc.
Then, the master equation (\ref{eq:master}) can be rewritten for positive and negative scaled overlaps, $\xi'>0$ and $\xi'<0$, as
\begin{eqnarray}
\frac{\partial}{\partial\phi}P_\phi(\xi') &=&
\int_0^\infty\left[T_{CC}(\xi'|\xi)P_\phi(\xi)-T_{CC}(\xi|\xi')P_\phi(\xi')\right]d\xi\nonumber\\
&+& \int_{-\infty}^0\left[T_{VC}(\xi'|\xi)P_\phi(\xi)-T_{CV}(\xi|\xi')P_\phi(\xi')\right]d\xi~,\nonumber\\
\label{eq:master_p}\\
\frac{\partial}{\partial\phi}P_\phi(\xi') &=&
\int_{-\infty}^0\left[T_{VV}(\xi'|\xi)P_\phi(\xi)-T_{VV}(\xi|\xi')P_\phi(\xi')\right]d\xi\nonumber\\
&+&\int_0^\infty\left[T_{CV}(\xi'|\xi)P_\phi(\xi)-T_{VC}(\xi|\xi')P_\phi(\xi')\right]d\xi~,\nonumber\\
\label{eq:master_n}
\end{eqnarray}
respectively.
The second terms on the right-hand-sides of Eqs.\ (\ref{eq:master_p}) and (\ref{eq:master_n}) are \emph{cross-terms of positive and negative overlaps} due to closing and opening contacts.
The other form of the master equation (\ref{eq:master_gamma}) can be rewritten as well.
In Appendix \ref{app:derive}, we derive the master equation, Eqs.\ (\ref{eq:master_p}) and (\ref{eq:master_n}), from the Chapman-Kolmogorov equation (\ref{eq:chapman-kolmogorov}).

\begin{figure}
\includegraphics[width=\columnwidth]{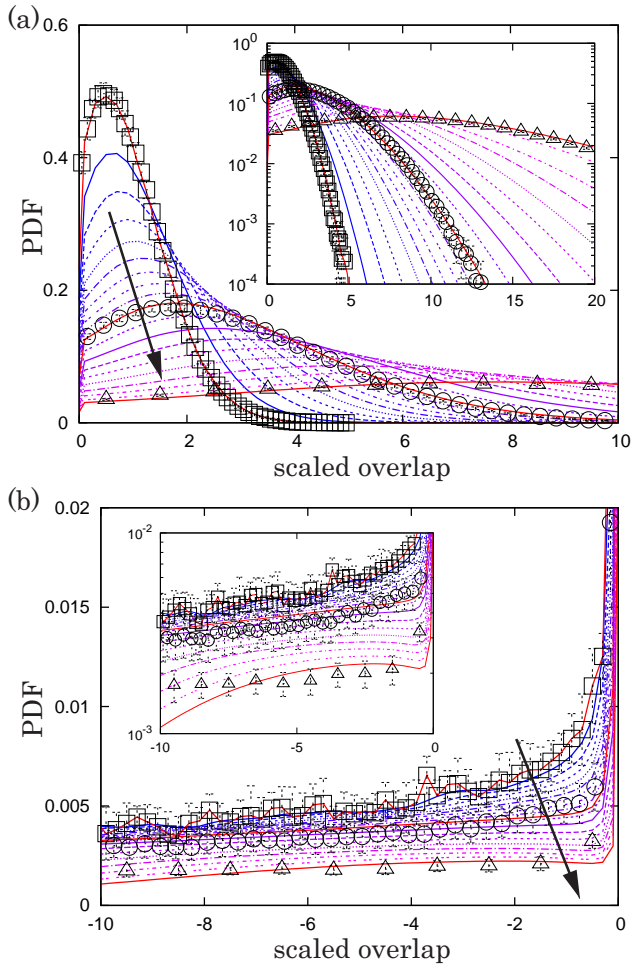}
\caption{(Color online)
Comparisons between the master equation (solid and dashed lines) and MD simulations (open symbols) under compression in (a) positive and (b) negative overlaps,
where the open squares, circles, and triangles are the PDFs obtained from our MD simulations at $\phi-\phi_J=4\times10^{-3}$, $1.2\times10^{-2}$, and $4\times10^{-2}$, respectively.
Overlaps are scaled by the averaged overlap at $\phi_0-\phi_J=4\times10^{-3}$ and the numerical solutions develop in the directions indicated by the arrows.
The insets show the corresponding semilogarithmic plots.
\label{fig:mrkv}}
\end{figure}

In our framework, transitions between overlaps are assumed to be Markov processes.
To validate this assumption, we compare numerical solutions of the master equation with the PDFs obtained from MD simulations.
Figure \ref{fig:mrkv} displays the numerical solutions under compression,
where the initial PDF is given by the MD simulation with the distance from jamming, $\phi_0-\phi_J=4\times10^{-3}$.
In this figure, the overlaps are scaled by the averaged overlap at the initial state, $\bar{x}(\phi_0)$,
where the increment of area fraction is fixed to $\delta\phi=4\times10^{-5}$ such that the scaled strain increment is less than $\gamma=\delta\gamma_0\equiv\delta\phi/(\phi_0-\phi_J)=10^{-2}$.
We confirm a good agreement between the numerical solutions (red solid lines) and MD simulations (open symbols) even in the tails of the PDFs
(the insets in Fig.\ \ref{fig:mrkv}), where the sequential solutions between $4\times10^{-3}\le\phi-\phi_J\le4\times10^{-2}$ are also displayed (dashed lines).

\begin{figure}
\includegraphics[width=6cm]{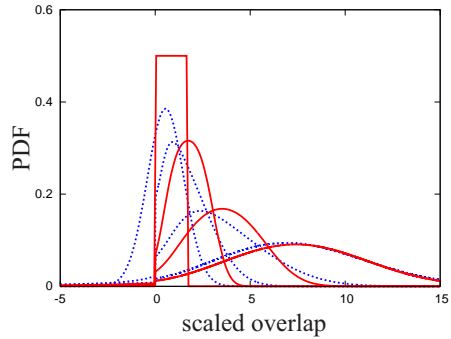}
\caption{(Color online)
The dependence of the master equation on the initial conditions, where the red solid and blue dotted lines
are numerical solutions (under compression) starting with a step function and Gaussian distribution, respectively.
\label{fig:imrk}}
\end{figure}

Note that the master equation becomes insensitive to the initial condition, which is the only input in our framework, after finite strain steps.
Figure \ref{fig:imrk} shows numerical solutions under compression with different initial conditions,\ i.e.\ a step function and Gaussian distribution,
where the solutions for both initial conditions converge to the same PDF after some increments of area fraction.
%

\begin{figure}
\includegraphics[width=8cm]{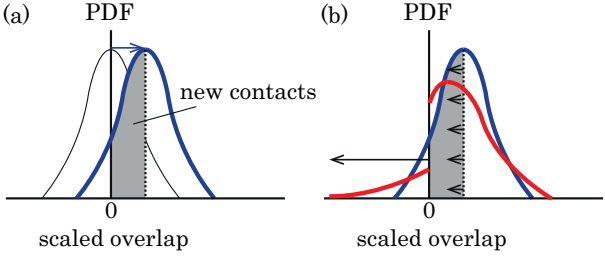}
\caption{(Color online)
Schematic pictures of (a) affine and (b) non-affine changes of a PDF.
(a) The affine deformation shifts the initial Gaussian PDF (thin solid line) to the new PDF (blue solid line)
as indicated by the arrow, where the shaded area corresponds to the amount of new contacts.
(b) The PDF after affine deformation (blue solid line) changes to the PDF after non-affine deformation (red solid line)
as indicated by arrows, where the typical penetration lengths of new contacts and virtual contacts are significantly different, $\lambda_c\ll\lambda_v$.
\label{fig:generating_gaps}}
\end{figure}

The master equation generates (or maintains) discontinuous ``jumps" of the PDFs around zero-overlap as we have observed after non-affine deformations in Fig.\ \ref{fig:pdfs}.
Figure \ref{fig:imrk_gaus} displays numerical solutions of the master equation, where the initial PDF is given by a Gaussian distribution which is initially continuous around zero.
As can be seen, a discontinuous jump is generated after several increments of area fraction.
Figure \ref{fig:generating_gaps} explains that a discontinuous jump is caused by the significant difference
between the dimensionless lengths in the CPDs in (CV) and (VC),\ i.e.\ $\lambda_c\ll\lambda_v$:
When virtual contacts are closed by affine deformation as indicated by the shaded area in Fig.\ \ref{fig:generating_gaps}(a) (no contact is opened by affine compression),
each new contact is pushed back by the repulsive interparticle force (the short arrows in Fig.\ \ref{fig:generating_gaps}(b))
so that the typical penetration is reduced to $\lambda_c$, which is considerably smaller than the maximum penetration by affine deformation, $B_a$.
Then, the overlaps between new contacts decrease and the PDF near zero-overlap (on the positive side) grows.
However, once particles are detached from each other, generated negative overlaps can increase freely
since there is no interparticle force and the typical magnitude of generated negative overlaps, $\lambda_v$, is much larger than the typical penetration, $\lambda_c$.
Thus, generated negative overlaps widely distribute such that the PDF near zero-overlap on the negative side becomes smaller than that on the positive side.
As a result, a jump between positive and negative sides is generated (or maintained).
Note that such discontinuities in the PDFs are specific to static packings which will disappear once a finite temperature is imposed \cite{th0,th1,th2}.
%
\begin{figure}
\caption{(Color online)
The generation of a discontinuous jump by the master equation, where the initial PDF (blue open squares) is a Gaussian distribution.
Numerical solutions of the master equation (red open circles) evolve from (a) to (d) under compression.
\label{fig:imrk_gaus}}
\end{figure}
%
\subsection{Irreversible responses}
\label{sub:macro}
%
We next turn to irreversible mechanical responses of soft particle packings to quasi-static deformations.

\begin{figure*}
\includegraphics[width=\textwidth]{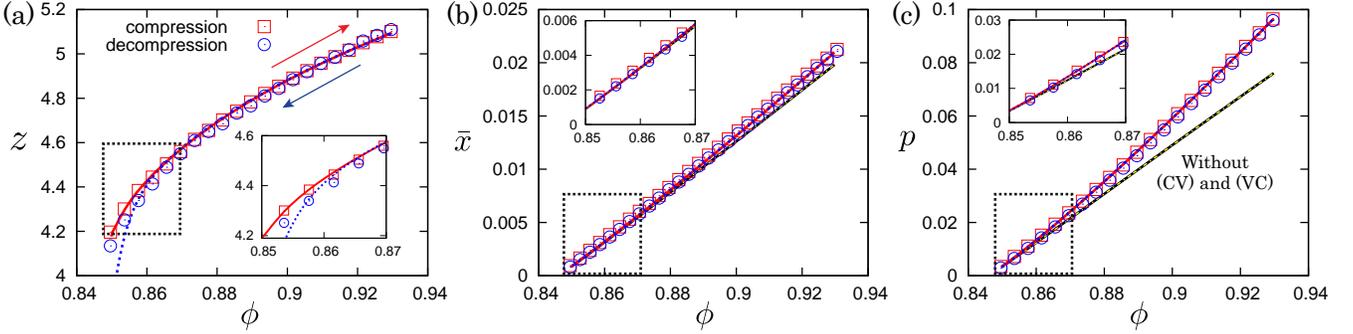}
\caption{(Color online)
(a) Coordination number, $z$, (b) averaged overlap, $\bar{x}$, and (c) pressure, $p$, in units of the mean diameter, $\bar{\sigma}$, and spring constant, $k$,
plotted against the area fraction during a compression-decompression cycle (the arrows in (a)).
The open symbols and lines are obtained from MD simulations and the master equation, respectively.
The (red) squares and (red) solid lines are the results under compression, while the (blue) circles and (blue) dotted lines are the results under decompression.
The straight lines in (b) and (c) are given by numerical solutions of the master equation without any opening and closing contacts,\ i.e.\ $W_{CV}=W_{VC}=0$,
where the (black) solid and (yellow) dotted lines are the results under compression and decompression, respectively.
The insets are zooms into the squares surrounded by the broken lines.
\label{fig:momt}}
\end{figure*}

Figure \ref{fig:momt} shows (a) the coordination number, (b) averaged overlap, and (c) static pressure under compression and decompression,
where we first increase the area fraction from $\phi_0-\phi_J=4\times10^{-3}$ to $\phi_1-\phi_J=8\times10^{-2}$
and then decrease it to $\phi_0-\phi_J$ with the increment, $\delta\phi=4\times10^{-4}$, assuming that $\phi_J$ is not changed by the cyclic loading \cite{nishant}.
In this figure, the lines are calculated by Eqs.\ (\ref{eq:momt_z})-(\ref{eq:momt_p}),
where we substituted numerical solutions of the master equation for the PDF in the $n$-th moment, Eq.\ (\ref{eq:moment}).
The open symbols are the results of MD simulations, where a reasonable agreement with the master equation is established.
In addition, the master equation captures irreversible responses of the macroscopic quantities (the coordination number in Fig.\ \ref{fig:momt}(a) is more visible)
and well reproduces the non-linear behavior of the static pressure (Fig.\ \ref{fig:momt}(c)) \cite{gn2}.
Note that the master equation \emph{without any opening and closing contacts},\ i.e.\ numerical solutions with zero transition rates in (CV) and (VC),
gives a linear increase and decrease of pressure (straight lines in Figs.\ \ref{fig:momt}(b) and (c))
as described in some literature focusing on the systems close to jamming \cite{gn3}.
\subsubsection{Irreversibility of the PDFs}
\begin{figure}
\includegraphics[width=\columnwidth]{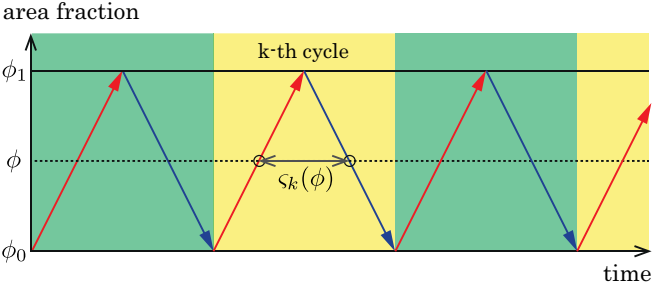}
\caption{(Color online)
A schematic picture of cyclic (de)compressions:
During compression, the area fraction increases from $\phi_0$ to $\phi_1$ along the red arrows,
while it decreases from $\phi_1$ to $\phi_0$ along the blue arrows during decompression.
The difference, $\varsigma_k(\phi)$, is defined in the $k$-th cycle at the same area fraction, $\phi$,
where the PDFs (calculated at the open circles) are compared with each other.
\label{fig:cyclic_scheme}}
\end{figure}

To quantify the degree of irreversibility, we introduce a difference between the PDFs during compression and decompression for the same area fraction,
$P^\mathrm{c}_{\phi,k}(\xi)$ and $P^\mathrm{d}_{\phi,k}(\xi)$, as
\begin{equation}
\varsigma_k(\phi)\equiv\sqrt{\left\langle\left[P^\mathrm{c}_{\phi,k}(\xi)-P^\mathrm{d}_{\phi,k}(\xi)\right]^2\right\rangle_\xi}~.
\label{eq:sigk}
\end{equation}
Here, the subscript, $k=1,2,\dots$, represents the number of cyclic (de)compressions
and the superscripts, $\mathrm{c}$ and $\mathrm{d}$, mean the PDFs during compression and decompression, respectively.
On the right-hand-side of Eq.\ (\ref{eq:sigk}), the bracket, $\langle\dots\rangle_\xi$, represents the average over all scaled overlaps ($-\infty<\xi<\infty$).
Figure \ref{fig:cyclic_scheme} displays a schematic picture of cyclic (de)compressions,
where the difference, $\varsigma_k(\phi)$, is calculated in the $k$-th cycle at the same area fraction, $\phi$.

\begin{figure*}
\includegraphics[width=\textwidth]{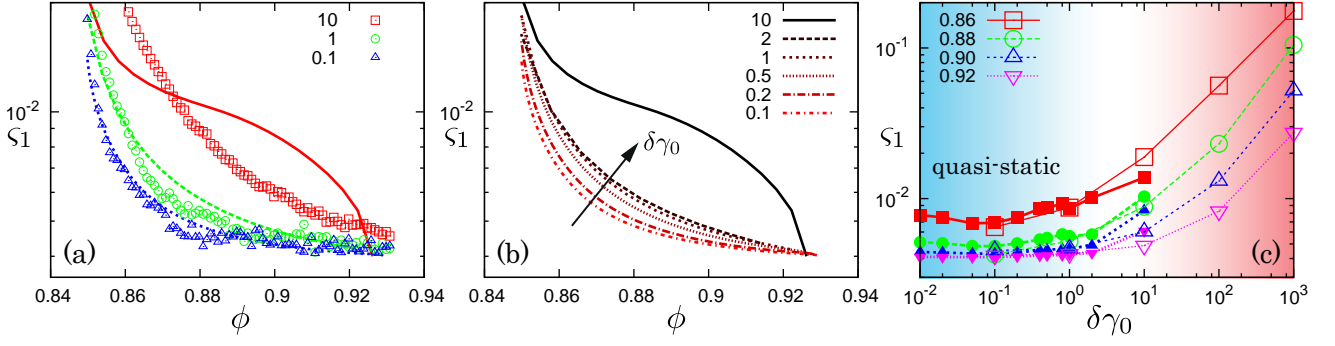}
\caption{(Color online)
(a) The difference for the first cycle, $\varsigma_1(\phi)$, plotted against the area fraction,
where the open symbols and lines are obtained from MD simulations and the master equation, respectively.
Different colors represent different strain increments defined at the initial state,\ i.e.\ $\delta\gamma_0\equiv\delta\phi/(\phi_0-\phi_J)=0.1$, $1$, and $10$, as listed in the legend.
(b) The difference, $\varsigma_1(\phi)$, obtained from the master equation, where the scaled strain increment, $\delta\gamma_0$, increases as indicated by the arrow and listed in the legend.
(c) The dependence of $\varsigma_1(\phi)$ on $\delta\gamma_0$, where the open and closed symbols are the results of MD simulations and the master equation, respectively.
Different symbols represent different area fractions as listed in the legend.
\label{fig:sig1}}
\end{figure*}

First, we study the dependence of irreversibility on the scaled strain increment.
As shown in Fig.\ \ref{fig:sig1}(a), the difference in the first cycle, $\varsigma_1(\phi)$, is a decreasing function of the area fraction,
where a good agreement between the MD simulation (the open triangles) and the master equation (the blue dotted line) with a small strain increment, $\delta\gamma_0=0.1$, is established.
Both the differences, $\varsigma_1(\phi)$, obtained from MD simulations and the master equation increase with the scaled strain increment,
where the master equation starts to deviate from MD simulations once $\delta\gamma_0$ exceeds unity (Figs.\ \ref{fig:sig1}(a) and (b)).
Figure \ref{fig:sig1}(c) shows the dependence of $\varsigma_1(\phi)$ on $\delta\gamma_0$,
where the irreversibilities observed in MD simulations (open symbols) are well reproduced by the master equation (closed symbols)
if the deformation is quasi-static (blue-shaded region, $\delta\gamma_0\lesssim0.1$).
Remarkably, the difference between loading and unloading remains finite, which implies that a strong irreversibility present even if the applied strain is very small.

\begin{figure*}
\includegraphics[width=\textwidth]{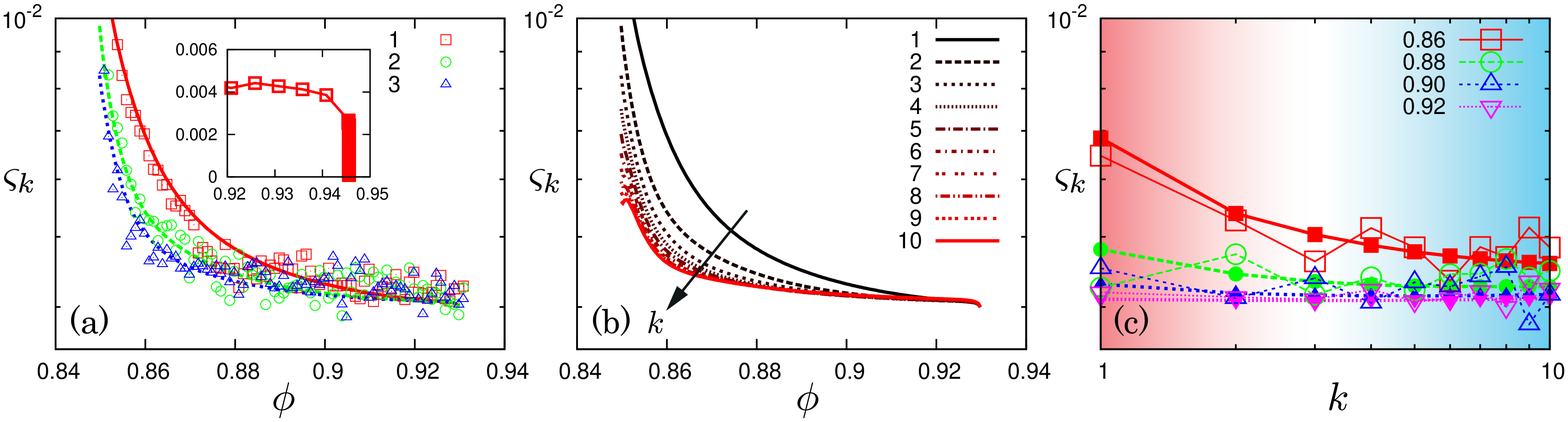}
\caption{(Color online)
(a) The difference for the $k$-th cycle, $\varsigma_k(\phi)$, plotted against the area fraction,
where we use $\delta\gamma_0=0.1$ and the open symbols and lines are obtained from MD simulations and the master equation, respectively.
Different colors represent the different number of cycles, $k$, as listed in the legend.
The inset shows a zoom-in around the maximum area fraction, $\phi_1$.
(b) The difference, $\varsigma_k(\phi)$, obtained from the master equation with $\delta\gamma_0=0.1$, where the number of cycles, $k$, increases as indicated by the arrow and listed in the legend.
(c) The dependence of $\varsigma_k(\phi)$ on $k$, where the open and closed symbols are the results of MD simulations and the master equation, respectively.
Different symbols represent different area fractions as listed in the legend.
\label{fig:sigk}}
\end{figure*}

Second, we examine the dependence of irreversibility on the number of cyclic (de)compressions.
Figure \ref{fig:sigk}(a) displays the difference, $\varsigma_k(\phi)$, for the different number of cycles, $k$,
where good agreements between MD simulations (the open symbols) and the master equation (the lines) are established with $\delta\gamma_0=0.1$.
(In the inset of Fig.\ \ref{fig:sigk}(a), we confirm that the difference decays to zero at the turning point, $\phi=\phi_1$.)
In this figure, the difference always decreases with the number of cycles (Figs.\ \ref{fig:sigk}(a) and (b)),
where the irreversibilities observed in MD simulations are well predicted by the master equation (Fig.\ \ref{fig:sigk}(c)).
%
%
\subsubsection{Irreversibility and local symmetry}
The microscopic origin of irreversible responses can be explained by our observations of the transition rates:
If the system response is reversible against deformations,
the transition rates for decompression and compression must satisfy a local symmetry,\ i.e.\ $T_{-\gamma}(\xi'|\xi)=T_\gamma(\xi|\xi')$.
This means that the scattered data of scaled overlaps for compression and decompression for the same $\gamma$ (Figs.\ \ref{fig:emsc_com} and \ref{fig:emsc_dec})
are symmetric with respect to the diagonal line, $\xi'=\xi$.
If we assume such a symmetry, however, the mean values of $\xi'$ and standard deviations under decompression should be given by
$m_l^\ast(\xi)=\xi/(a_l+1)-b_l/(a_l+1)$ and $v_l^\ast=v_l/(a_l+1)$, respectively, which is clearly different from our results, $m_l(\xi)=(1-a_l)\xi-b_l$ and $v_l$.
Note that $m_l^\ast(\xi)$ and $v_l^\ast$ converge to $m_l(\xi)$ and $v_l$ in the small strain limit,\
i.e.\ $m_l^\ast(\xi)=(1-a_l)\xi-b_l+O(\gamma^2)\simeq m_l(\xi)$ and $v_l^\ast=v_l+O(\gamma^2)\simeq v_l$,
so that the scattered data tend to be almost symmetric around the diagonal line if $\gamma\ll1$.
On the other hand, the transition rates in (CV) and (VC) are finite even when the scaled strain increment is very small,
which leads the finite irreversibility during cyclic (de)compressions.
In fact, the number of opening and closing contacts obeys a power law, $\mathcal{E}_{CV}, \mathcal{E}_{VC}\sim\gamma^{0.7}$, as observed in Fig.\ \ref{fig:delsum}.
From these observations, the irreversible responses cannot be avoided if the applied strain is finite (not zero).
\subsubsection{Asymptotic behavior near jamming}
We also examine asymptotic behaviors of the PDFs near the jamming transition, $\phi\rightarrow\phi_J$, by the master equation.
Here, we numerically solve the master equation by fixing the scaled increment, where the increment of area fraction, $\delta\phi=(\phi-\phi_J)\gamma$,
decreases as the system approaches to the jamming transition such that the system is always above jamming.
Then, we introduced a difference between the PDFs, $P_{\phi+\delta\phi}(\xi)$ and $P_\phi(\xi)$, as
\begin{equation}
\chi(\phi) = \sqrt{\left\langle\left[P_{\phi+\delta\phi}(\xi)-P_\phi(\xi)\right]^2\right\rangle_\xi}~,
\label{eq:PphJ}
\end{equation}
where we find that the difference, $\chi(\phi)$, asymptotically decreases to zero (Fig.\ \ref{fig:phJ_gfix}),
implying that the PDF tends to be \emph{self-similar} near jamming.
Note that, however, the decay is so slow, $\chi(\phi)\sim(\phi-\phi_J)^{0.14}$, and thus it is practically impossible to reach the asymptotic limit.

\begin{figure}
\caption{(Color online)
The asymptotic decrease of the difference between the PDFs, $\chi(\phi)$, plotted against the distance from jamming, $\phi-\phi_J$ (the open circles).
Here, we fixed the scaled strain increment to $\gamma=0.1$,\ i.e.\ increments of area fraction are in the range, $10^{-12}\le\delta\phi\le10^{-4}$.
The dotted line represents a power law fitting, $\chi(\phi)=7.6\times10^{-4}(\phi-\phi_J)^{0.14}$.
Note that the initial data drop around $\phi-\phi_J\simeq10^{-3}$ (the open circles in the right-top)
is caused by the fluctuations of the initial PDF, which is given by the MD simulation.
\label{fig:phJ_gfix}}
\end{figure}
%
\section{Discussion and Conclusion}
\label{sec:discon}
In summary, we have proposed and numerically investigated a master equation for the probability distribution functions (PDFs) of forces
in soft particle packings under isotropic compression and decompression.

First, mechanical responses of soft particle packings to quasi-static deformations are determined by the restructuring of force-chain networks,
involving their complicated recombinations of force-chains,\ i.e.\ opening and closing contacts.
To take into account all kinds of changes of contacts, we have introduced Delaunay triangulations of soft particle packings,
where the force-chain networks are generalized to include not only the particles in contact, but also the nearest neighbors in \emph{virtual contact}.
Then, the statistics of structuring of force-chain networks are well captured by the PDFs of generalized overlaps,
where the ``overlap" (or interparticle gaps) between the particles in virtual contact is defined as negative values.
In addition, the rates of four kinds of changes of contacts,\ i.e.\ contact-to-contact (CC), virtual-to-virtual (VV), contact-to-virtual (CV), and virtual-to-contact (VC),
are described by the conditional probability distributions (CPDs) in the Chapman-Kolmogorov equation (\ref{eq:chapman-kolmogorov}),
or equivalently, by the transition rates in the master equation (\ref{eq:master}).
We have numerically determined the transition rates for the four kinds of changes by MD simulations
of two-dimensional bidispersed frictionless soft particle packings under both isotropic compression and decompression.

Second, the transition rates are defined as distributions of scaled overlaps around their mean values,
where we have found that the mean values are described by linear functions of the scaled overlaps before deformation.
Then, the deviations from affine deformation, characterized by the excess slopes ($a_l$) and offsets ($b_l$),
linearly increase with the scaled strain increment, $\gamma=\delta\phi/(\phi-\phi_J)$, for both isotropic compression and decompression.
The scaled overlaps after non-affine deformation have finite widths ($v_l$) around their mean values, where the widths also linearly increase with $\gamma$.
This is the evidence of stochastic processes of overlaps and interparticle gaps during the restructuring of force-chain networks.
All the linear scalings against $\gamma$ are not affected by the system size.
Note that affine deformations do not generate such a scatter so that their transition rates are given by delta-functions, resulting in a deterministic evolution of the PDFs.

Third, we have found that the transition rates for (CC) and (VV) are symmetric around the mean values
even though the PDFs are asymmetric and cannot be described by conventional distribution functions.
The transition rates can be unified to the $q$-Gaussian distributions,
where their tails are wider than normal Gaussian distributions implying that the stochastic evolution of scaled overlaps is correlated.
Note that the tail of the transition rate in (VV) is much wider than that in (CC),\ i.e.\
the correlation between interparticle gaps, interestingly, seems to be much stronger than that between contacts.
We have also confirmed that the transition rates for (CV) and (VC) are given by exponential distributions satisfying normalization conditions for $\xi>0$ and $\xi<0$, respectively.
In the transition rates for (CV) and (VC), the dimensionless length scales ($\lambda_l$) represent typical penetration- and gap-lengths after creating contacts and virtual contacts, respectively,
where their significant difference ($\lambda_c\ll\lambda_v$) induces the discontinuous jump of the PDF around zero-overlap.

Fourth, we have validated the master equation, in which stochastics of scaled overlaps are assumed to be Markov processes and the initial PDF is the only input,
by comparing numerical solutions of the master equation with the PDFs obtained from MD simulations.
We have confirmed that the evolution of the PDFs is well described by the master equation and their dependence on the initial condition vanishes with increasing strain (deformation).
We have also demonstrated the generation of a discontinuous jump of the PDF by numerically solving the master equation.

Finally, irreversible responses of macroscopic quantities,\ e.g.\ coordination number, averaged overlap, and static pressure,
defined by the moments of scaled overlaps are well reproduced by the master equation.
Introducing the difference between the PDFs under cyclic compressions and decompressions for the same area fraction,
we have confirmed that the degree of irreversibility observed in MD simulations is well reproduced by the master equation if the system undergoes quasi-static deformations.
In both MD simulations and the master equation, we have found that the difference is a decreasing function of the area fraction and the number of cyclic deformations.
We have also confirmed the self-similarity of the PDFs by the master equation near the jamming transition,
where the difference between the PDFs asymptotically decays to zero as the area fraction approaches to the jamming area fraction, $\phi_J$.

In our recent study \cite{saitoh3}, we have confirmed the basic properties of the master equation,\
i.e.\ linear scalings for the excess slopes, offsets, and fluctuations against $\gamma$, and the symmetry of transition rates,
by MD simulations of two-dimensional polydisperse frictional particles, where we have found that the increase of polydispersity and decrease of friction
broaden the tails of transition rates, implying the increase of correlations of contacts and interparticle gaps.
In addition, the transition rates have been found to be symmetric $q$-Gaussian distributions in our recent experiments of wooden cylinders \cite{saitoh4}.

Because the master equation requires the increment $\delta\phi$ to be much smaller than $\phi-\phi_J$,\ i.e.\ $\gamma\ll1$, it can never reach $\phi_J$.
This means that the jamming transition is a singular limit of the master equation, but possibly available to asymptotic analysis.
In addition, there is the need of further studies on the physical origin of the stochastics of overlaps described above.
Analytic or asymptotic solutions of the master equation are also important steps towards the understanding of functional forms of the PDFs.
Our analysis can be easily extended to three dimensions and the extension to other cases is also straightforward,\ e.g.\
the solutions under shear can be obtained if we apply our results for compression and decompression to principal compressive and tensile directions, respectively \cite{saitoh5}.
%
%
\begin{acknowledgments}
We thank M. Sperl, L. E. Silbert, B. P. Tighe, H. Hayakawa, S. Takesue, S. Yukawa, T. Hatano,
H. Yoshino, S. Inagaki, T. Mitsudo, K. Kanazawa, G.\ Combe, V.\ Richefeu, and G.\ Viggiani for fruitful discussions.
This work was financially supported by the NWO-STW VICI grant 10828 and a part of numerical
computation in this work was carried out at the Yukawa Institute Computer Facility, Kyoto, Japan.
\end{acknowledgments}

\appendix
\section{Discretizations of the PDFs, CPDs, and the Chapman-Kolmogorov equation}
\label{app:pdf}
In this Appendix, we explain how to calculate the PDFs and CPDs from the numerical data.

We first make a histogram of scaled overlaps before (de)compression as $\mathrm{hist}[\xi]$.
Here, we count the number of particle pairs, $i$ and $j$, of which scaled overlaps, $\xi_{ij}$, are in between
\begin{equation}
\xi-\Delta\xi/2\le\xi_{ij}\le\xi+\Delta\xi/2~,
\label{eq:binning_xi}
\end{equation}
where $\Delta\xi$ is a small increment or \emph{bin-size} for $\xi$.
Then, the PDF of scaled overlaps is defined as
\begin{equation}
P_\phi(\xi) = \frac{\mathrm{hist}[\xi]}{N_p\Delta\xi}~,
\label{eq:cal_pdf}
\end{equation}
where $N_p$ is the number of Delaunay edges (the total number of contacts and virtual contacts) which is given by
\begin{equation}
N_p = \sum_{-\infty\le\xi\le\infty}\mathrm{hist}[\xi]~.
\label{eq:Np}
\end{equation}
From Eqs.\ (\ref{eq:cal_pdf}) and (\ref{eq:Np}), the normalization condition for the PDF,\ Eq.\ (\ref{eq:norm_pdf}),\ is automatically satisfied as
\begin{equation}
\sum_{-\infty\le\xi\le\infty}\frac{\mathrm{hist}[\xi]}{N_p\Delta\xi}\Delta\xi=1~.
\end{equation}

After (de)compression, we make another histogram of new scaled overlaps as $\mathrm{hist}[\xi'|\xi]$.
Here, we count the number of particle pairs, $i$ and $j$, of which scaled overlaps after (de)compression, $\xi'_{ij}$, are in between
\begin{equation}
\xi'-\Delta\xi'/2\le\xi'_{ij}\le\xi'+\Delta'\xi/2~,
\end{equation}
where $\Delta\xi'=\Delta\xi$ is a bin-size for $\xi'$.
Note that $\xi'_{ij}$ were $\xi_{ij}$ which satisfied Eq.\ (\ref{eq:binning_xi}) before (de)compression.
The new histogram, $\mathrm{hist}[\xi'|\xi]$, is connected with $\mathrm{hist}[\xi]$ through the following relations:
\begin{eqnarray}
\mathrm{hist}[\xi] &=& \sum_{-\infty\le\xi'\le\infty}\mathrm{hist}[\xi'|\xi]~, \label{eq:nor_hist1} \\
\mathrm{hist}[\xi'] &=& \sum_{-\infty\le\xi\le\infty}\mathrm{hist}[\xi'|\xi]~. \label{eq:nor_hist2}
\end{eqnarray}
Then, the joint probability distribution is introduced as
\begin{equation}
P(\xi';\xi) = \frac{\mathrm{hist}[\xi'|\xi]}{N_p\Delta\xi\Delta\xi'}~.
\label{eq:cal_joint}
\end{equation}

The CPD is defined as the ratio of the joint probability distribution to the PDF \cite{vanKampen}.
Therefore, the CPD is given by using Eqs.\ (\ref{eq:cal_pdf}) and (\ref{eq:cal_joint}) as
\begin{eqnarray}
W(\xi'|\xi)
&=& \frac{P(\xi';\xi)}{P_\phi(\xi)} \nonumber\\
&=& \frac{\mathrm{hist}[\xi'|\xi]}{\mathrm{hist}[\xi]\Delta\xi'}~.
\label{eq:cal_cpd}
\end{eqnarray}
It is readily found that Eq.\ (\ref{eq:cal_cpd}) satisfies the normalization condition for the CPD,\ Eq.\ (\ref{eq:norm_cpd}),\ i.e.\
\begin{equation}
\sum_{-\infty\le\xi'\le\infty}
\frac{\mathrm{hist}[\xi'|\xi]}{\mathrm{hist}[\xi]\Delta\xi'}\Delta\xi'=1~,
\end{equation}
where we used Eq.\ (\ref{eq:nor_hist1}).

From Eqs.\ (\ref{eq:cal_pdf}) and (\ref{eq:cal_cpd}), the Chapman-Kolmogorov equation is derived as
\begin{eqnarray}
\int_{-\infty}^\infty P_\phi(\xi)W(\xi'|\xi)d\xi &\simeq&
\sum_{-\infty\le\xi\le\infty} \frac{\mathrm{hist}[\xi]}{N_p\Delta\xi}
\frac{\mathrm{hist}[\xi'|\xi]}{\mathrm{hist}[\xi]\Delta\xi'}\Delta\xi \nonumber\\
&=& \frac{1}{N_p\Delta\xi'} \sum_{-\infty\le\xi\le\infty}\mathrm{hist}[\xi'|\xi] \nonumber\\
&=& \frac{\mathrm{hist}[\xi']}{N_p\Delta\xi'}~,
\label{eq:cal_chapman}
\end{eqnarray}
which yields the discretized PDF after (de)compression, $P_{\phi\pm\delta\phi}(\xi')$.
%
\section{A derivation of the master equation}
\label{app:derive}
In this Appendix, we derive complete forms of the master equation,\ i.e.\ Eqs.\ (\ref{eq:master_p}) and (\ref{eq:master_n}),
from the Chapman-Kolmogorov equation (\ref{eq:chapman-kolmogorov}).

At first, we divide the Chapman-Kolmogorov equation (\ref{eq:chapman-kolmogorov}) for \emph{positive} scaled overlaps as
\begin{eqnarray}
P_{\phi+\delta\phi}(\xi') &=& \int_0^\infty W_{CC}(\xi'|\xi)P_\phi(\xi)d\xi \nonumber\\
& & + \int_{-\infty}^0 W_{VC}(\xi'|\xi)P_\phi(\xi)d\xi~,
\label{app:eq:ck_p}
\end{eqnarray}
where the first and second terms on the right-hand-side represent gains of contacts from \emph{contacts} (CC) and \emph{virtual contacts} (VC), respectively.
Then, the difference between the PDFs after and before non-affine deformation is given by
\begin{eqnarray}
& & P_{\phi+\delta\phi}(\xi')-P_\phi(\xi') = \nonumber\\
& & \int_0^\infty W_{CC}(\xi'|\xi)P_\phi(\xi)d\xi + \int_{-\infty}^0 W_{VC}(\xi'|\xi)P_\phi(\xi)d\xi\nonumber\\
& &-P_\phi(\xi')\left\{\int_0^\infty W_{CC}(\xi|\xi')d\xi + \int_{-\infty}^0 W_{CV}(\xi|\xi')d\xi\right\}\nonumber\\
& &=\int_0^\infty\left\{W_{CC}(\xi'|\xi)P_\phi(\xi)-W_{CC}(\xi|\xi')P_\phi(\xi')\right\}d\xi\nonumber\\
& &+\int_{-\infty}^0\left\{W_{VC}(\xi'|\xi)P_\phi(\xi)-W_{CV}(\xi|\xi')P_\phi(\xi')\right\}d\xi,
\label{app:eq:ck_p2}
\end{eqnarray}
where we used the normalization condition of the CPDs, Eq.\ (\ref{eq:norm_pos}).
Dividing Eq.\ (\ref{app:eq:ck_p2}) by a small increment of area fraction, $\delta\phi$, we find the master equation for positive scaled overlaps ($\xi'>0$) as
\begin{eqnarray}
\frac{\partial}{\partial\phi}P_\phi(\xi') &=&
\int_0^\infty\left[T_{CC}(\xi'|\xi)P_\phi(\xi)-T_{CC}(\xi|\xi')P_\phi(\xi')\right]d\xi\nonumber\\
&+& \int_{-\infty}^0\left[T_{VC}(\xi'|\xi)P_\phi(\xi)-T_{CV}(\xi|\xi')P_\phi(\xi')\right]d\xi~,\nonumber\\
\label{app:eq:master_p}
\end{eqnarray}
where the transition rates in (CC), (VC), and (CV) are defined as
\begin{eqnarray}
T_{CC}(\xi'|\xi) &\equiv& \lim_{\delta\phi\rightarrow0}\frac{W_{CC}(\xi'|\xi)}{\delta\phi}~, \label{app:eq:trate(CC)}\\
T_{VC}(\xi'|\xi) &\equiv& \lim_{\delta\phi\rightarrow0}\frac{W_{VC}(\xi'|\xi)}{\delta\phi}~, \label{app:eq:trate(VC)}\\
T_{CV}(\xi'|\xi) &\equiv& \lim_{\delta\phi\rightarrow0}\frac{W_{CV}(\xi'|\xi)}{\delta\phi}~, \label{app:eq:trate(CV)}
\end{eqnarray}
respectively.

Similarly, we divide the Chapman-Kolmogorov equation for \emph{negative} scaled overlaps as
\begin{eqnarray}
P_{\phi+\delta\phi}(\xi') &=& \int_{-\infty}^0 W_{VV}(\xi'|\xi)P_\phi(\xi)d\xi \nonumber\\
& & + \int_0^\infty W_{CV}(\xi'|\xi)P_\phi(\xi)d\xi~,
\label{app:eq:ck_n}
\end{eqnarray}
where the first and second terms on the right-hand-side represent gains of virtual contacts from \emph{virtual contacts} (VV) and \emph{contacts} (CV), respectively.
The difference between the PDFs after and before non-affine deformation is given by
\begin{eqnarray}
& & P_{\phi+\delta\phi}(\xi')-P_\phi(\xi') = \nonumber\\
& & \int_{-\infty}^0 W_{VV}(\xi'|\xi)P_\phi(\xi)d\xi + \int_0^\infty W_{CV}(\xi'|\xi)P_\phi(\xi)d\xi\nonumber\\
& &-P_\phi(\xi')\left\{\int_{-\infty}^0 W_{VV}(\xi|\xi')d\xi + \int_0^\infty W_{VC}(\xi|\xi')d\xi\right\}\nonumber\\
& &=\int_{-\infty}^0\left\{W_{VV}(\xi'|\xi)P_\phi(\xi)-W_{VV}(\xi|\xi')P_\phi(\xi')\right\}d\xi\nonumber\\
& &+\int_0^\infty\left\{W_{CV}(\xi'|\xi)P_\phi(\xi)-W_{VC}(\xi|\xi')P_\phi(\xi')\right\}d\xi,
\label{app:eq:ck_n2}
\end{eqnarray}
where we used the normalization condition of the CPDs, Eq.\ (\ref{eq:norm_neg}).
Dividing Eq.\ (\ref{app:eq:ck_n2}) by $\delta\phi$, we find the master equation for negative scaled overlaps ($\xi'<0$) as
\begin{eqnarray}
\frac{\partial}{\partial\phi}P_\phi(\xi') &=&
\int_{-\infty}^0\left[T_{VV}(\xi'|\xi)P_\phi(\xi)-T_{VV}(\xi|\xi')P_\phi(\xi')\right]d\xi\nonumber\\
&+&\int_0^\infty\left[T_{CV}(\xi'|\xi)P_\phi(\xi)-T_{VC}(\xi|\xi')P_\phi(\xi')\right]d\xi~,\nonumber\\
\label{app:eq:master_n}
\end{eqnarray}
where the transition rate in (VV) is defined as
\begin{equation}
T_{VV}(\xi'|\xi) \equiv \lim_{\delta\phi\rightarrow0}\frac{W_{VV}(\xi'|\xi)}{\delta\phi}~.
\label{app:eq:trate(VV)}
\end{equation}
\bibliography{master}
\end{document}